\documentclass[review]{elsarticle}
\usepackage{array}
\usepackage{amssymb}
\usepackage{lineno}
\usepackage{hyperref}
\usepackage{setspace}
\usepackage{subfigure}
\usepackage{float}
\usepackage{graphicx}

\modulolinenumbers[5]

\journal{Ad Hoc Networks}

\begin{document}
\begin{frontmatter}

\title{Impact of Human Behavior on Social Opportunistic Forwarding}


\author[mymainaddress]{Waldir Moreira\corref{mycorrespondingauthor}}
\ead{waldir.junior@ulusofona.pt}

\cortext[mycorrespondingauthor]{This is the author's pre-print version. Personal use of this material is permitted. However, permission to reprint/republish this material for advertising or promotion or for creating new collective works for resale or for redistribution to thirds must be obtained from the copyright owner. The camera-ready version of this work is being published at Elsevier Ad Hoc Networks, 2014 and is property of Elsevier B.V.}

\author[mymainaddress]{Paulo Mendes}

\address[mymainaddress]{COPELABS, University Lusofona, Lisbon, Portugal}

\begin{abstract}
The current Internet design is not capable to support communications
in environments characterized by very long delays and frequent network
partitions. To allow devices to communicate in such environments,
delay-tolerant networking solutions have been proposed by exploiting
opportunistic message forwarding, with limited expectations of end-to-end
connectivity and node resources. Such solutions envision non-traditional
communication scenarios, such as disaster areas and development regions.
Several forwarding algorithms have been investigated, aiming to offer
the best trade-off between cost (number of message replicas) and rate
of successful message delivery. Among such proposals, there has been
an effort to employ social similarity inferred from user mobility
patterns in opportunistic routing solutions to improve forwarding.
However, these research effort presents two major limitations: first,
it is focused on distribution of the intercontact time over the complete
network structure, ignoring the impact that human behavior has on
the dynamics of the network; and second, most of the proposed solutions
look at challenging networking environments where networks have low
density, ignoring the potential use of delay-tolerant networking to
support low cost communications in networks with higher density, such
as urban scenarios. This paper presents a study of the impact that
human behavior has on opportunistic forwarding. Our goal is twofold:
i) to show that performance in low and high density networks can be
improved by taking the dynamics of the network into account; and ii)
to show that the delay-tolerant networking can be used to reduce communication
costs in networks with higher density by taking the behavior of the
user into account.\end{abstract}
\end{frontmatter}


\section{Introduction}

Wireless devices have become more portable and with increased capabilities
(e.g., processing, storage), which is creating the foundations for
the deployment of pervasive wireless networks, encompassing personal
devices (e.g. smartphones and tablets). Additionally, wireless technology
has been extended to allow direct communication: vehicle-to-vehicle
- for safety information exchange; device-to-device - aiming at 3G
offloading; Wi-Fi direct - overcome the need for infrastructure entities
(i.e., access points).

The combination of pervasive wireless devices and direct wireless
communication solutions can be used to support the deployment of two
major type of applications: end-to-end communication in development
regions, since today's Internet routing protocols may operate poorly
in such environments, characterized by very long delay paths and frequent
network partitions; and low cost communication, namely data sharing,
in urban scenarios, to bypass expensive data mobile communications
and the unreliable presence of open Wi-Fi access points. 

These networking scenarios (from development regional to large urban
scenarios) are characterized by network graphs with different densities,
which pose different challenges in terms of data forwarding. The challenge
that we aim to tackle in this paper is to investigate the impact of
human behavior on opportunistic forwarding, namely the awareness about
users' social and data similarities.

Most of the prior art has been studying data transfer opportunities
between wireless devices carried by humans, by looking at the distribution
of the intercontact time, which is the time gap separating two contacts
between the same pair of devices \cite{impactHuman}. In challenging
networking environments, opportunistic contacts among mobile devices
may improve communications among peers as well as content dissemination,
mitigating the effects of network disruption. This gave rise to the
investigation of opportunistic networks, of which Delay-Tolerant Networks
(DTN) are an example, encompassing different forwarding proposals
to quickly send data from one point to another even in the absence
of an end-to-end path between them. Such proposals range from flooding
content \cite{epidemic} in the network up to solutions that take
into account the social interactions among users \cite{socialcast,contentPlace,bubble2011,dlife,cipro,scorp2}.
In the latter case, wireless contacts are aggregated into a social
graph, and a variety of metrics (e.g., centrality and similarity)
or algorithms (e.g., community detection) have been proposed to assess
the utility of a node to deliver a content or bring it closer to the
destination. Nevertheless, the structure of such graphs is rather
dynamic, since users\textquoteright{} social behavior and interactions
vary throughout their daily routines. This brings us to our first
assumption: forwarding algorithms should be able to exploit social
graphs that reflect people\textquoteright{}s dynamic behavior. Prior
art have studied forwarding algorithms that consider only the global
network structure, without taking people\textquoteright{}s behavior
into account \cite{thyneighbor}. In this paper, we show that forwarding
algorithms that exploit social graphs reflecting the variations in
people\textquoteright{}s daily routines are able to  improve the
performance of social-aware opportunistic networking.

Our second challenge was to analyze how to expand the deployment of
DTN technology, which is normally seen only as useful to allow communications
in challenging environments, such as development regions. For this
study, we focus on data sharing since this should be the most interesting
application to take advantage of low cost communication in dense networks,
such as in an urban scenario. In this case, we studied two hypothesis
to ensure good performance when the density of the network increases:
i) forwarding based on social graphs, where aggregation is based only
on social similarities; and ii) forwarding based on behavior graphs,
where aggregation is done by combining different aspects of human
behavior, such as social similarities and data similarities (derived
from the interests that users demonstrate in specific type of data). 

Hence, in this paper we aim to investigate the possibility of developing
an opportunistic forwarding system able to support low-cost services
in dense networking scenarios as well as basic services in extreme
networking conditions, by exploiting social as well as data similarities
among users. Our work shows which type of opportunistic forwarding
scheme is more suitable for delay-tolerant applications, based on
the density of the network in scenarios spanning from developing regions
to urban environments. Our findings lead to a new research challenge
aiming to expand the impact of DTNs: the investigation of self-awareness
mechanisms able to adapt their forwarding schemes based on the context
of the user, namely the density of the network where he/she is currently.

The remainder of the paper is structured as follows. Section \ref{sec:Motivation}
aims to motivate our work, namely in what concerns the goal to study
methods to expand the deployment of DTNs, and the impact that a better
understanding of human behavior can have in the development of efficient
forwarding solutions. In Section \ref{sec:Network-Scenario-Characterizatio}
we present our definition of network density based on the deployment
scenarios that we look at to pursuit our study and experiments. Section
\ref{sec:OppForwarding} presents a set of forwarding algorithms that
are considered in our study, including our proposals. In Section \ref{sec:Experiments},
we show the performance results of opportunistic forwarding over different
network densities. Section \ref{sec:Conclusions} concludes our work,
and identifies future research challenges to expand the impact of
DTNs.

\section{Motivation \label{sec:Motivation}}

The growing number of mobile devices equipped with a wireless interface
and the end-user trend to shift toward wireless technology are opening
new possibilities for networking. In particular, opportunistic communication
embodies a feasible solution for environments with scarce or costly
infrastructure-based connectivity. A lot of attention has been given
to the development of opportunistic forwarding solutions for networks
with scarce connectivity, which are considered a natural fit for DTN
technology. However, it is our belief that opportunistic forwarding
can also be applied to more dense networks, were Internet communications
are expensive, or applications aim to take advantage of direct communications
among people.

In what concerns challenging networks, the most common approach has
been to make use of social similarities to improve performance over
the overall network. In this case, our work is motivated by the fact
that such approaches ignore the behavior patterns that people present
in their daily routines \cite{behaviour}, which may lead to further
performance improvements.

In what concerns the application of DTN and opportunistic forwarding
to dense networks, most of the prior art aims to implement a store-carry-and-forward
communication model that exploits specific devices found in urban
scenarios, such as buses \cite{carpool} and cars \cite{geoDTNnav}.
It is our understanding that the development of opportunistic forwarding
solutions should not depend upon specific equipments only found is
some scenarios, since this mitigates the deployment expansion of such
proposals. It is our belief that the success of the DTN technology
depends on its deployment range, which can only be ensured if such
technology is based on pervasive wireless devices, such as smartphones:
these devices are present in development regions as well as urban
scenarios. In the latter case, communications between smartphones
can also exploit mobility patterns of different vehicles ridden by
people.

In order to design useful applications, it is vital to have a good
understanding of the target environment and its users. Different types
of user behavior may result in different network conditions and shall
have a huge impact on whether or not a particular application is of
interest to the user. A fair amount of work has been done on studying
human mobility traces in order to gain understanding of real life
mobility patterns and how those affect the properties of the opportunistic
networks that are possible in that environment \cite{PocketSwitchedNetworksHumanMobilityConferenceEnvironments2005,impactHuman,ComplexNetworkAnalysisHumanMobility}.
Although mobility patterns are important properties of the network,
it is also important to understand the impact the human behavior,
such as data interests, have on these networks. Hence, our work aims
to tackle this new research trend, expanding social awareness to human
behavior awareness. Among the different human behavior metrics that
can be considered, we focus our attention on data similarities since
data sharing is the most common application in the Internet. The study
of data similarities depends on which applications are in place in
the network and how the users use them. Usage patterns also depend
on the users' context, so the same data patterns do not apply to all
users. Approximations of some use cases might be possible to derive
from the way cellular networks are used, but that will most likely
not be applicable to all types of applications.

Looking at data similarities may improve the performance of opportunistic
forwarding \cite{socialcast,contentPlace}. However it is not clear
if the improvement is higher than exploiting social interactions and
structure (i.e., communities \cite{bubble2011}, as well as levels
of social interaction \cite{dlife,cipro}). Thus, combining social
and data similarities shall bring benefits (i.e., faster, better content
reachability) to opportunistic forwarding.\emph{ }Hence, in this work
we aim to show when the exploitation of social similarities results
in a good performance, and when such performance can be augmented
by combining them with data similarity metrics.

\section{Network Scenario Characterization\label{sec:Network-Scenario-Characterizatio}}

One can observe that a networking scenario may vary according to its
density. Sparse scenarios are characterized by very long delays (e.g.,
space communications \cite{farrell2006delay}) and communication
suffers with frequent disruption mostly due to the lack of infrastructure
and geographic location (e.g., rural areas \cite{oracleknowledge},
riverside communities \cite{socialDTN,DigitalInclusion}). It is
common to see solutions relying on message ferries \cite{msgferry}
or data mules \cite{datamules} as to overcome the missing infrastructure.
This is a classic scenario whose challenges are more related to transport
protocols (i.e., dealing with extremely high delays) than routing
itself.

In what concerns dense scenarios, communication can take place through
both infrastructured (e.g., access points, cell towers) and infrastructureless
(e.g., WiFi direct, bluetooth) means. Still, disruption remains a
problem, but now seen from a different perspective: the dynamic bebavior
of users (e.g., high mobility), different sources of interference
(e.g., overlapping spectrum), poor coverage (e.g., areas full of closed
access points) are factors that may contribute to link intermittency,
despite all the available surrounding infrastructure. With the advances
in the industry for portable devices and wireless technologies, this
type of scenario is easily observed nowadays in urban settings.

In this work, we see dense (urban) scenarios as imposing new research
challenges for opportunistic networks given the aforementioned characteristics.
Due to the popularity and capabilities of mobile devices, users want
to be able to sent and retrieve data anytime and anywhere. In other
words, a free Internet scenario with low cost denominator networking
imposed to users is a reality.

With this in mind, we define the density of the network according
to the surroundings of the users and independently of the existence
of infrastructure: what matters is i) that nodes can communicate directly;
and ii) that the average node degree reflects the number of contact
opportunities a node may have in such specific scenarios. We studied
the characteristics of the three scenarios which are considered in
the experiments in Sec. \ref{sec:Experiments}: the CRAWDAD traces
of Cambridge \cite{cambridge-haggle-imote-content-2006-09-15} that
corresponds to contacts of 36 students during their daily activities,
the MIT \cite{mit-reality-2005-07-01} traces comprising 97 Nokia
6600 smart phones distributed among the students and staff of this
institution, and the synthetic mobility scenario that encompasses
150 walking people, following the \emph{Shortest Path Map Based Movement}
model (i.e., nodes randomly choose destinations and use the shortest
path to reach them). With the Gephi v0.8.2 \cite{gephi} analysis
tool, we accounted for the network densities of these scenarios summarized
in Table \ref{tab:Network-densities}.

\begin{table}
\begin{centering}
\caption{\label{tab:Network-densities}Network densities of the considered
scenarios}

\par\end{centering}

\centering{}%
\begin{tabular}{|c|c|c|c|}
\hline 
\textbf{Scenario} & \textbf{Cambridge} & \textbf{MIT} & \textbf{Synthetic}\tabularnewline
\hline 
\textbf{Identified density} & 26.83 & 47.01 & 148.80\tabularnewline
\hline 
\end{tabular}
\end{table}

Table \ref{tab:Network-densities} displays the scenarios in increasing
order of density. Thus, it is expected that the routing solutions
have an increasing performance behavior as network density increases.
This is due to the different contact opportunities that a node may
have, which increase (and therefore can be beneficial) for routing
purposes.

\section{Opportunistic Forwarding in Wireless Networks\label{sec:OppForwarding}}

This section presents the most relevant and latest opportunistic forwarding
proposals, considering whether they make use of social and/or data
similarity metrics. Similarity metrics are used to build graphs over
which such forwarding proposals operate \cite{bookchapter}. That
is, instead of considering the number and frequency of contacts due
to the mobility of hosts, such approaches take into account more stable
social (e.g., common social groups and communities, node popularity,
levels of centrality, social relationships and interactions, user
profiles) and/or data (e.g., shared interests, interest of users in
the content traversing network, content availability, type of content)
aspects, aiming to reduce the cost of opportunistic forwarding. Moreover,
opportunistic forwarding proposals may take into account the dynamics
of user behavior, i.e., the resulting social graphs may consider what
happens in terms of social interactions throughout the daily routine
of the users.

Table \ref{tab:Opportunistic-Forwarding} summarizes the type of similarity
(i.e., social and/or data) considered by the opportunistic forwarding
proposals and whether (or not) they consider the observed user behavior
to build dynamic social graphs.

\begin{table}
\caption{\label{tab:Opportunistic-Forwarding}Opportunistic Forwarding Proposals}

\begin{centering}
\begin{tabular}{|>{\centering}p{2.2cm}|>{\centering}p{3cm}|>{\centering}p{3.5cm}|>{\centering}p{1.7cm}|}
\hline 
\textbf{Proposals} & \textbf{Social similarity metrics} & \textbf{Data similarity metrics} & \textbf{Dynamic graphs}\tabularnewline
\hline 
Bubble Rap & Communities and centrality &  & \tabularnewline
\hline 
CiPRO & User profile &  & \tabularnewline
\hline 
SocialCast &  & Shared interests & \tabularnewline
\hline 
ContentPlace & Social relationship and communities & Interest on the content & \tabularnewline
\hline 
dLife & Social weight and node importance &  & \checkmark\tabularnewline
\hline 
SCORP & Social weight & Content type and interest on the content & \checkmark\tabularnewline
\hline 
\end{tabular}
\par\end{centering}

\end{table}

\textit{Bubble Rap }\cite{bubble2011}, \emph{CiPRO} \cite{cipro},
\textit{SocialCast} \cite{socialcast}, and\emph{ ContentPlace} \cite{contentPlace}
belong to the category that considers social similarity and/or data
similarity metrics, but does not suitably reflect the dynamism of
user behavior in the underlying social graph. 

\textit{Bubble Rap} combines node centrality with the notion of community
to make forwarding decisions. The centrality metric identifies hub
nodes inside (i.e., local) or outside (i.e., global) communities.
Messages are replicated based on global centrality until they reach
the community of the destination host (i.e., a node belonging to the
same community). Then, it uses the local centrality to reach the destination
inside the community.

\emph{CiPRO} considers the time and place nodes meet throughout their
routines.\emph{ CiPRO }holds knowledge of nodes (e.g., carrier's name,
address, nationality, ...) expressed by means of profiles that are
used to compute the encounter probability among nodes in specific
time periods. Nodes that meet occasionally get a copy of the message
only if they have higher encounter probability towards its destination.
If nodes meet frequently, history of encounters is used to predict
encounter probabilities for efficient broadcasting of control packets
and messages.

\textit{SocialCast} considers the interest shared among nodes. It
devises a utility function that captures the future co-location of
the node (with others sharing the same interest) and the change in
its connectivity degree. Thus, the utility function measures how good
message carrier a node can be regarding a given interest. \textit{SocialCast}
functions are based on the publish-subscribe paradigm, where users
broadcast their interests, and content is disseminated to interested
parties and/or to new carriers with high utility.

\emph{ContentPlace} considers information about the users' social
relationships to improve content availability. It computes a utility
function for each data object considering: i) the access probability
to each object and the involved cost in accessing it; ii) the social
strength of the user towards the different communities which he/she
belongs to and/or has interacted with. The idea is having the users
to fetch data objects that maximize the utility function with respect
to local cache limitations, and choosing those objects that are of
interest to users and can be further disseminated in the communities
they have strong social ties.

The next category considers solely social similarity metrics and take
into account the dynamism of user behavior while building the underlying
social graph. \emph{dLife} \cite{dlife} is in this category and
it takes into account the dynamism of users' behavior found in their
daily life routines to aid forwarding. The goal is to keep track of
the different levels of social interactions (in terms of contact duration)
nodes have throughout their daily activities in order to infer how
well socially connected they are in different periods of the day.
Forwarding takes place by considering either the social strength (i.e.,
weight) among users or their importance in specific time periods.

Finally, the last category comprises both social and data similarity
metrics and the dynamism observed in the behavior of users.\emph{
SCORP} \cite{scorp2} belongs to this category. It considers the
type of content and the social relationship between the parties interested
in such content type. \emph{SCORP} nodes are expected to receive and
store messages considering their own interests as well as interests
of other nodes with whom they have interacted before. Data forwarding
takes place by considering the social weight of the encountered node
towards nodes interested in the message that is about to be replicated.

For the remainder of this paper, we consider one representative from
each of the described categories: \emph{Bubble Rap}, for being solely
based on social similarity metrics; \emph{dLife} and \emph{SCORP},
for considering social and data similarity metrics and for being the
proposals which satisfactory capture the dynamic user behavior\emph{
}in the resulting social graphs. These proposals are enough to help
us illustrate how the performance of opportunistic routing proposals
in networks with different densities can be further improved by considering
the user dynamic behavior.

\section{Evaluation of Opportunistic Forwarding over Different Network Density
Scenarios\label{sec:Experiments}}

In this section, we analyze the performance behavior of\emph{ Bubble
Rap}, \emph{dLife,} and \emph{SCORP} over different network density
scenarios. With the experiments in this section, we want i) to show
that performance in low (Sec. \ref{sub:LowDensityNetwork}) and high
(Sec. \ref{sub:HighDensityNetwork}) density networks can be improved
by taking the dynamics of the network into account; and ii) to show
that the delay-tolerant networking can be used to reduce communication
costs in networks with higher density by taking the behavior of the
user into account (Sec. \ref{sub:DifferentDensities}).

\subsection{Methodology and Simulation Settings\label{sub:Evaluation-Methodology}}

Simulations are carried out in the Opportunistic Network Environment
(ONE) simulator \cite{one}. Results are presented with a 95\% confidence
interval and in terms of averaged delivery probability (i.e., ratio
between the number of delivered messages and the number of messages
that should have been delivered), cost (i.e., number of replicas per
delivered message), and latency (i.e., time elapsed between message
creation and delivery).

The trace scenarios comprise 36 (Cambridge) and 97 (MIT) nodes carrying
devices during their daily activities. The synthetic mobility scenario
simulates 3 groups ($A$, $M$, and $B$) of 50 people each, who carry
nodes equipped with 250-Kbps Bluetooth interfaces, and moving with
speed up to 1.4 m/s. The reason for considering traces and synthetic
mobility scenario relates to the fact that: i) with the former, we
have a representation of real user behavior; and ii) with the latter,
we are able to have a network density much higher from the perspective
of the user, as defined in Sec. \ref{sec:Network-Scenario-Characterizatio}.
Most analysis have been based on datasets with low density, collected
in a constrained setting, which is not representative for realistic
use cases of the networks being studied. If one is interested in the
properties of a large scale urban environment, it is probably not
meaningful to study traces collected from 36 or 97 user at a conference
or university campus.

Across all experiments, proposals experience the same load and number
of messages that must reach the destinations. In the Cambridge trace
(cf. Sec. \ref{sub:LowDensityNetwork}), the \emph{Bubble Rap}/\emph{dLife}
source sends 1, 5, 10, 20 and 35 different messages to each of the
35 destinations, while the \emph{SCORP} source creates 35 messages
with unique content types, and the receivers are configured with 1,
5, 10, 20, and 35 randomly assigned interests. Thus, we have a total
of 35, 175, 350, 700, and 1225 generated messages. The \emph{msg/int}
notation represents the number of messages sent by \emph{Bubble Rap}
and \emph{dLife} sources, or the number of interests of each of the
\emph{SCORP} receivers. Since \emph{Bubble Rap}/\emph{dLife} sources
generate more messages, in this scenario node $0$ (the source) has
no buffer restriction and message generation varies with the load:
35 messages/day rate (load of 1, 5, and 10 messages), and 70 and 140
messages/day rates (load of 20 and 35 messages, respectively).

As for the synthetic mobility scenario (cf. Sec. \ref{sub:HighDensityNetwork}),
200 messages are generated. With \emph{Bubble} \emph{Rap }and \emph{dLife},
node $0$ (group \emph{$A$}) generates 100 messages to nodes in groups
$B$ and $M$, and node $100$ (group $B$) generates 100 messages
to nodes in groups $A$ and $M$. For \emph{SCORP}, each group has
different interests: group $A$ (\emph{reading}), group $B$ (\emph{games}),
and group $M$ (\emph{reading} and \emph{games}). The source nodes,
$0$ and $100$, generate only one message for each content type,
\emph{game} and \emph{reading}. This guarantees the same number of
messages expected to be received, i.e., 200. Also, by varying the
node pause times between 100 and 100000 seconds, we have different
levels of mobility (varying from 3456 to 3.4 movements in the simulation).
In this scenario, all source nodes have restricted buffer, but rate
is of 25 messages every 12 hours. This is done so that \emph{Bubble
Rap}/\emph{dLife} do not discard messages prior to even trying exchange/deliver
them given the buffer constraint.

Finally, in Sec. \ref{sub:DifferentDensities}, the load generated
is equivalent to 6000, 78000, and 200 messages to be delivered across
all experiments for Cambridge, MIT, and synthetic mobility scenarios,
respectively. 

Regarding message TTL, we set it to be unlimited in order to observe
the performance behavior (i.e., buffer consumption, number of replicas)
of the forwarding proposals in networks with high traffic load. Message
size ranges from 1 to 100 kB. Despite nodes may have plenty of storage,
we consider nodes having different capabilities (i.e., smartphones).
Thus, nodes have buffers limited to 2 MB as we consider that nodes
may not be willing to share all their storage space. The performance
evaluation follows the guidelines of a Universal Evaluation Framework
(UEF) \cite{latincom,ieeeLA} to guarantee fairness in the assessment.

As for proposals, \emph{Bubble Rap }uses the K-Clique and cumulative
window algorithms for community formation and centrality computation
as in \cite{bubble2011}. As for \emph{dLife }and \emph{SCORP}, both
consider 24 daily samples (i.e., each of one hour) as mentioned in
\cite{scorp2}.

\subsection{Performance over Low Density Network\label{sub:LowDensityNetwork}}

\begin{figure}[h]
\centering{}\includegraphics[scale=0.7]{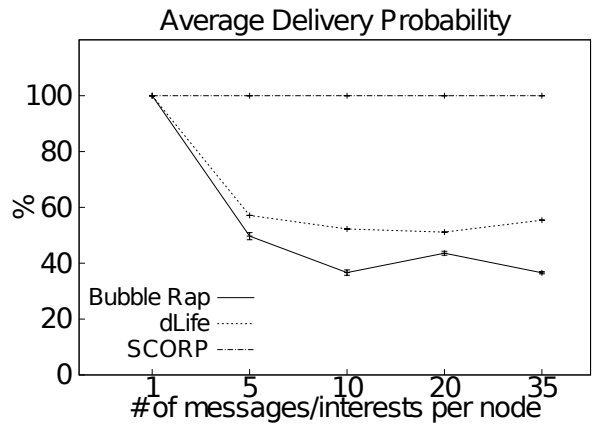}\caption{\label{fig:DeliveryLoad}Delivery under different network loads}
\end{figure}

This section presents the performance of opportunistic forwarding
proposals over a low density network scenario. Fig. \ref{fig:DeliveryLoad}
presents the average delivery probability with different messages
and interests being generated.

In the 1 msg/int configuration, formed communities comprise almost
all nodes. This means that each node has high probability to meet
any other node, which is advantageous for \emph{Bubble Rap} since
most of its deliveries happen to nodes sharing communities. Due to
the dense properties of the network, \emph{dLife} and \emph{SCORP
}take advantage of direct delivery: 57\% and 51\% of messages, respectively,
are delivered directly to destinations.

As load increases, \emph{Bubble Rap }has an 50\% decrease in delivery
performance. This occurs since it relies on communities to perform
forwardings, and consequently buffer space becomes an issue. To support
this claim, we estimate buffer usage for the 5 msg/int configuration:
there is an average of 80340.7 forwardings, and if this number is
divided by the number of days (12%
\footnote{~In simulation it is worth \textasciitilde{}12 days of communications.%
}) and by the number of nodes (35, source not included), we get an
average of 191.28 replications per node. Multiplied by the average
message size (52kB), the buffer occupancy is roughly 9.94 MB in each
node, which exceeds\textbf{ }the 2MB allowed (cf. Sec. 4.1).

This estimation is for a worst case scenario, where \emph{Bubble Rap}
spreads copies to every encountered node. Since this cannot happen,
as \emph{Bubble Rap} also relies on local centrality to reduce replication,
buffer exhaustion is really an issue given that messages are replicated
to fewer nodes and not to all as in our estimation. As more messages
are generated, replication increases: this causes the spread of messages
that potentially take over forwarding opportunities from other messages,
reducing \emph{Bubble Rap}'s delivery capability.

\emph{dLife} has a 43\% performance decrease when network load increases,
as it takes time to have an accurate view of the social weights. This
leads to forwardings that never reach destinations given the contact
sporadicity. For the 10 msg/int configuration, \emph{dLife} also experiences
buffer exhaustion: estimated consumption is 2.17 MB per node. Still,
by considering social weights or node importance allows \emph{dLife}
a more stable behavior than \emph{Bubble Rap}.

Since content is only replicated to nodes that are interested in it,
or have a strong social interaction with other nodes interested in
such content, the delivery capability of SCORP raises as the ability
of nodes to become a good carrier increases (i.e., the more interests
a node has, the better it is to deliver content to others, since they
potentially share interests). The maximum estimated buffer consumption
of \emph{SCORP} is of 0.16 MB (35 msg/int).

\begin{figure}[h]
\centering{}\includegraphics[scale=0.7]{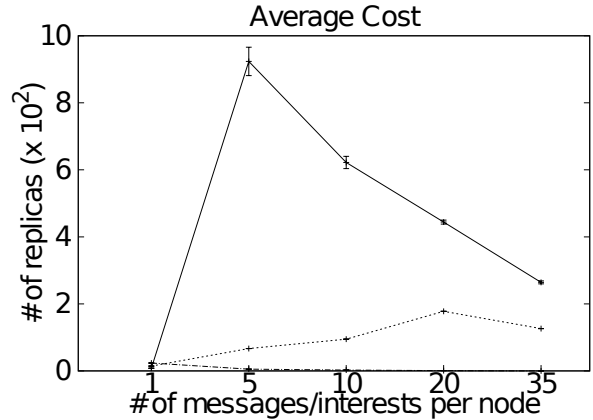}\caption{\label{fig:CostLoad}Cost under different network loads}
\end{figure}

Fig. \ref{fig:CostLoad} presents the average cost behavior. In the
1 msg/int configuration, all proposals create very few replicas to
perform a successful delivery, 7.95 (\emph{Bubble Rap}), 14.32 (\emph{dLife}),
and 23.46 (\emph{SCORP}), as they rely mostly on shared communities
and/or direct deliveries. We also observe that \emph{SCORP} produces
more replicas than \emph{dLife} due to a particularity in its implementation:
\emph{SCORP} nodes with interest in a specific content of a message
not only process it, but also replicate it to other interested nodes,
thus creating extra replicas.

For the 5, 10, 20 and 35 msg/int configurations, replication is directly
proportional to the load. Thus, cost is expected to increase as load
increases, as seen with \emph{dLife}. The same performance behavior
was expected for \emph{Bubble Rap}. However, the observed cost peaks
relate to the message creation time and contact sporadicity: when
a message is created in a period of high number of contacts, which
results in much more replications. This is more evident with \emph{Bubble
Rap} at the 5 msg/int configuration as it relies on shared communities
to forward: as mentioned earlier, most of the communities comprise
almost all nodes, which increases its replication rate.

Despite their efforts, these replications do not improve their delivery
probabilities, contributing only to the associated cost for performing
successful deliveries.

With more interests, a \emph{SCORP} node can serve as a carrier for
a larger number of nodes. Consequently, the observed extra replicas
make the proposal rather efficient: \emph{SCORP} creates an average
of 6.39 replicas across all msg/int configurations, while \emph{Bubble
Rap }and \emph{dLife} produce an average 452.41 and 96 replicas, respectively.

\begin{figure}[h]
\centering{}\includegraphics[scale=0.7]{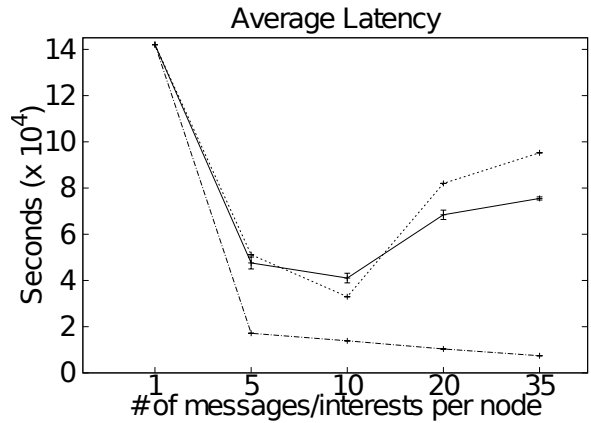}\caption{\label{fig:LatencyLoad}Latency under different network loads}
\end{figure}

Fig. \ref{fig:LatencyLoad} shows the average latency that messages
experience. The latency peak in the 1 msg/int configuration refers
to the message generation time: some messages are created during periods
where very few contacts (and sometimes none) take place followed by
long periods (12 to 23 hours) with almost no contact. Consequently,
messages are stored longer, contributing to the increase of the overall
latency. This effect is mitigated as the load increases with messages
being created almost immediately before a high number of contacts
take place.

Since latency is in function of the delivered messages, the decrease
and variable behavior of \emph{Bubble Rap }and \emph{dLife} is due
to their delivery rates decrease and increase, and also to their choices
of next forwarders that may take longer to deliver content to destinations.
\emph{SCORP} experiences latencies up to approx. 90.2\% and 92.2\%
less than \emph{Bubble Rap} and \emph{dLife}, respectively. The ability
of a node to deliver content increases with the number of its interests.
Thus, a node can receive more messages when it is interested in their
contents, and consequently becomes a better forwarder since the probability
of coming into contact with other nodes sharing similar interests
is very high, thus reducing latency.

\subsection{Performance over High Density Network\label{sub:HighDensityNetwork}}

This section presents the performance of opportunistic forwarding
proposals over a high density network scenario.

\begin{figure}[h]
\centering{}\includegraphics[scale=0.7]{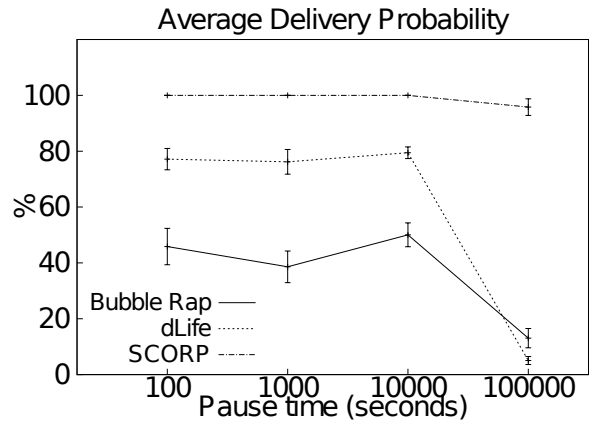}\caption{\label{fig:DeliveryMobility}Delivery under varied mobility rates}
\end{figure}

Fig. \ref{fig:DeliveryMobility} presents the average delivery probability.
Given the community formation characteristic of this scenario, \emph{Bubble
Rap} relies mostly on the global centrality to deliver content. By
looking at centrality \cite{bubble2011}, we observe very few nodes
(out of the 150) with global centrality that can actually aid in forwarding,
i.e., 19.33\% (29 nodes), 10.67\% (16 nodes), 21.33\% (32 nodes),
and 2\% (3 nodes) for 100, 1000, 10000, and 100000 pause time configurations,
respectively. So, these nodes become hubs and given buffer constraint
and infinite TTL (i.e., messages created earlier take the opportunity
of newly created ones), message drop is certain, directly impacting
\emph{Bubble Rap}.

Given the high number of contacts, the computation of social weight
and node importance done by \emph{dLife} takes longer to reflect reality:
thus \emph{dLife} replicates more and experiences buffer exhaustion.
Indeed, social awareness is advantageous, but still not enough to
reach optimal delivery rate in such conditions.

Independent of the number of contacts among nodes, \emph{SCORP} can
still identify nodes that are better related to others sharing similar
interests, reaching optimal delivery rate for 100, 1000, and 10000
pause time configurations. By considering nodes' interest in content
and their social weights, \emph{SCORP} does not suffer as much with
node mobility as \emph{dLife} and \emph{Bubble} \emph{Rap}.

With 100000 seconds of pause time, the little interaction happening
in a sporadic manner (with intervals between 20 and 26 hours) affects
\emph{Bubble Rap}, \emph{dLife }and \emph{SCORP} as they depend on
such interactions to compute centrality, node importance, and social
weights, as well as to exchange/deliver content.

\begin{figure}[h]
\centering{}\includegraphics[scale=0.7]{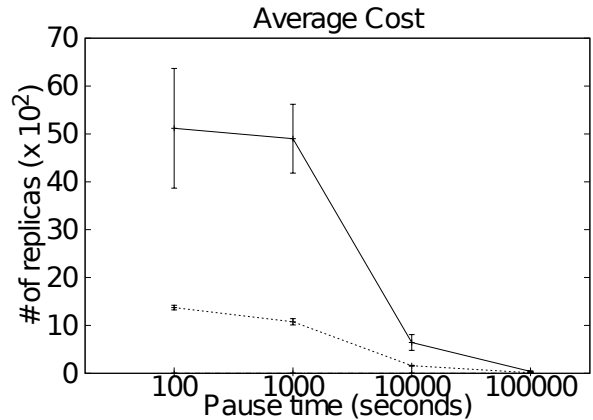}\caption{\label{fig:CostMobility}Cost under varied mobility rates}
\end{figure}

Fig. \ref{fig:CostMobility} presents the average cost behavior. As
pause time increases, the number of contacts among nodes decreases,
providing all solutions with the opportunity to have a stable view
of the network in terms of their social metrics with 100, 1000, and
10000 seconds of pause time. This explains the cost reduction experienced
by \emph{Bubble Rap} and \emph{dLife}: both are able to identify the
best next forwarders, which results in the creation of less replicas
to perform a successful delivery.

\emph{SCORP} has a very low replication rate (average of 0.5 replicas)
given its choice to replicate based on the interest that nodes have
on content and on their social weight towards other nodes interested
in such content. When the intermediate node has an increased number
of interests (i.e., by having different interests, the node can potentially
deliver more content) as observed in Sec. \ref{sub:LowDensityNetwork},
replication costs are even lower. Furthermore, \emph{SCORP }suitably
uses buffer space with an estimated average occupancy of 0.03 MB per
node per day.

With 100000 seconds of pause time, as cost is in function of delivered
messages (and deliveries are very low, due to contact sporadicity),
proposals have a low cost.

\begin{figure}[h]
\centering{}\includegraphics[scale=0.7]{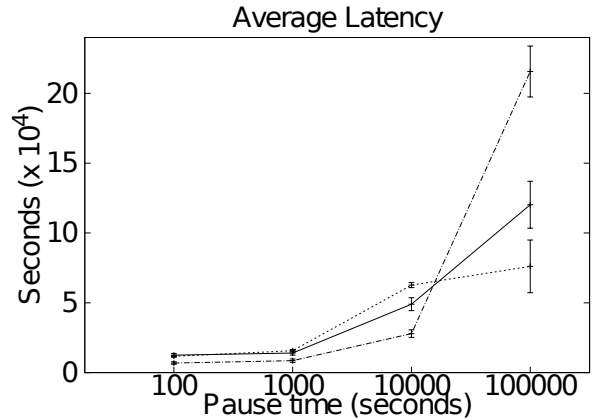}\caption{\label{fig:LatencyMobility}Latency under varied mobility rates}
\end{figure}

As expected (cf. Fig. \ref{fig:LatencyMobility}), latency increases
as node mobility decreases: encounters are less frequent, and so content
must be stored for longer times. Also, the time that the social metrics
take to converge (i.e., a more stable view of the network in terms
of centrality, social weight, and node importance) contributes for
the increase in the experienced latency. The highest increase in latency
with 100000 seconds of pause time is due to contacts happening in
a sporadic fashion with intervals between them of up to 26 hours,
thus proposals take much longer to perform a delivery.

\subsection{Performance over Different Network Densities\label{sub:DifferentDensities}}

\begin{figure}[h]
\centering{}\includegraphics[scale=0.7]{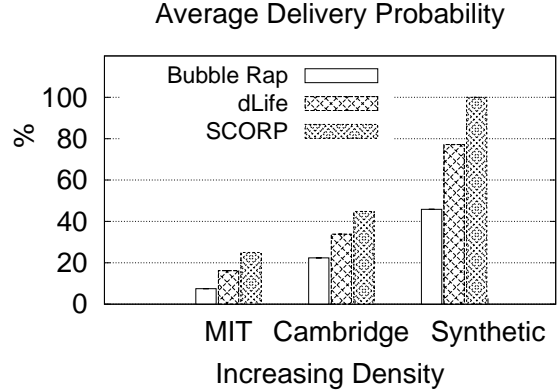}\caption{\label{fig:DeliveryDensity}Delivery under different network densities}
\end{figure}

This section shows how network density impacts on the performance
of opportunistic forwarding proposals. As mentioned before, we want
to bring attention to dense scenarios found in urban settings: a panoply
of heterogenous devices that could overcome disruption by interacting
directly with one another to improve the networking experience of
users. Fig. \ref{fig:DeliveryDensity} presents the average delivery
probability with different network densities.

As mentioned in Sec. \ref{sec:Network-Scenario-Characterizatio},
it was expected that the performance of social-aware opportunistic
routing improve with the increase of network density. However, we
can observe that content-oblivious \emph{Bubble Rap} and \emph{dLife}
experience a decrease in performance in the MIT scenario despite of
its identified density (47.01) being almost almost twice as the one
identified in Cambridge (26.83). The reason for such behavior lies
on the characteristics of each scenario, with MIT nodes covering a
much bigger area. Despite of having a higher number of contacts between
nodes, the MIT scenario may lead to messages reaching nodes that are
not the best forwarders, and these messages, given the unlimited TTL,
may end up taking the delivery opportunity of newly created messages.
This directly affects the performance of both \emph{Bubble Rap} and
\emph{dLife. }

Yet the content-oriented \emph{SCORP }overcomes such features of the
MIT scenario since it also considers those nodes that are interested
in the content being replicated or that are strongly related to interested
parties. Unlike the content-oblivious solutions, \emph{SCORP} has
a 6.14\% improvement despite the challenging scenario. 

Performance behavior for all proposal indeed improve with higher network
density (148.8, Synthetic scenario). The reason is tied to the fact
that a higher density indicates more contact opportunities for the
exchange of messages. 

\begin{figure}[h]
\centering{}\includegraphics[scale=0.7]{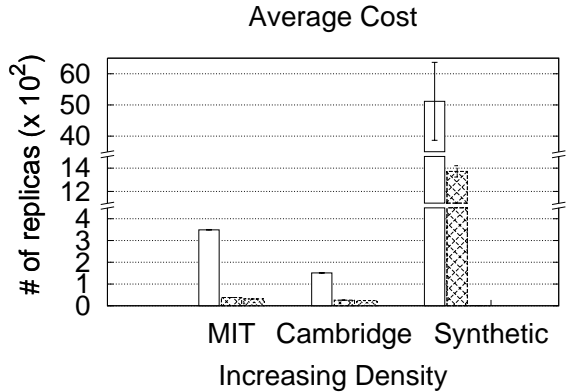}\caption{\label{fig:CostDensity}Cost under different network densities}
\end{figure}

Fig. \ref{fig:CostDensity} presents the average cost which is expected
to increase with network density. This is because the more contact
opportunities the scenario has, the more replicas are created by proposals.
This can be easily seen with \emph{Bubble Rap}, which creates an average
of more than 5000 replicas to perform a successful delivery. 

The same cost increasing trend is observed with \emph{dLife}, but
in different orders and especially for the synthetic scenario. We
believe that the much higher number of copies (up to 98\% when compared
to other scenarios) is related to the mobility rate. In the Synthetic
experiment nodes move a lot, which results in a high number of contacts
with many nodes. Consequently, \emph{dLife} takes longer to have a
stable view of the network in terms of its social weights and node
importances, which leads to the creation of unwanted replicas. \emph{SCORP}
has the best cost performance (up to 0.5 replicas to perform a successful
delivery), since the more interests a node has, the better forwarder
it is. The $M$ group (cf. Sec. \ref{sub:Evaluation-Methodology})
accounts for 50\% of the interests existing in the network, which
makes it a greater carrier for messages.

\begin{figure}[h]
\centering{}\includegraphics[scale=0.7]{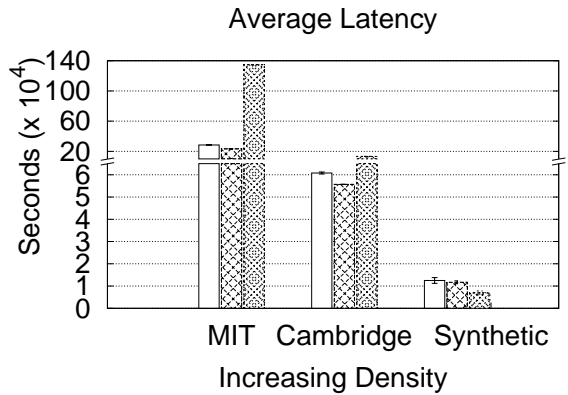}\caption{\label{fig:LatencyDensity}Latency under different network densities}
\end{figure}

Fig. \ref{fig:LatencyDensity} shows the average latency experienced
by messages since their creation up to their reception at destination.
In the traces experiments, all proposals keep the same trend: average
latency increases. This is due to the fact that nodes in these experiments
span different areas and the encounter frequency happens at different
rates. This can result in messages being forwarded to nodes who may
reach a given destination, but delivery time increases with the area
and duration of experiments. SCORP presents a much higher increase
in the MIT experiment, as it takes its time to suitably choose the
next forwarders (based on their interest on the message's content
or social relationship to other interested parties). This added to
fact that interactions among nodes happen according to area they move
jointly contribute to such latency peak.

As for the Synthetic mobility experiment, not only the proposals have
many different contact opportunities, but also nodes encounter more
frequently. This consequently has a positive impact for \emph{Bubble
Rap}, \emph{dLife}, and \emph{SCORP} that are able to deliver content
in less time (12486s, 11710s , and 6864s, respectively).

\section{Conclusions \label{sec:Conclusions}}

Opportunistic forwarding can aid communication in two major application
scenarios: end-to-end communication in development regions and low
cost communication in urban scenarios. Such scenarios do have different
network densities which adds more challenges to opportunistic forwarding.
The underlying graphs, over which these opportunistic forwarding proposals
operate, comprise (e.g., common social groups and communities, node
popularity, levels of centrality, social relationships and interactions,
user profiles) and/or data (e.g., shared interests, interest of users
in the content traversing network, content availability, type of content)
aspects. Additionally, such graphs may (or not) take into account
the dynamics of user behavior.

Thus, in this paper, we exploit the possibility of having an opportunistic
forwarding system that can provide support to i) low-cost services
in dense networking scenarios; and ii) basic services in extreme networking
conditions (e.g., communications in development regions), considering
social and data similarities among users as well as the dynamic behavior
found in the users' daily routines.

Our results show that opportunistic forwarding, based on social and
data similarity metrics and considering the dynamism observed in the
users behavior, does answer the communication needs of users in both
dense (i.e., urban) and challenged (i.e., development region) scenarios.
Performance improvements go up to 54\% regarding delivery capability
while latency and cost can be reduced by 45\% and 99\% respectively,
when compared to forwarding solely based on data similarity and completely
agnostic to user behavior.

These findings point to a new research challenge regarding the impact
of DTN application: the investigation of self-awareness mechanisms
able to adapt their forwarding schemes based on the context of the
user, namely the density of the network where he/she is currently.


\bibliographystyle{elsarticle-num}
\bibliography{bib-or}

\end{document}